\documentclass[letterpaper]{article} 
\usepackage[preprint]{aaai2027}  
\usepackage[hyphens]{url}  
\usepackage{graphicx} 
\usepackage{natbib}  
\usepackage{caption} 
\usepackage{algorithm}
\usepackage{algorithmic}

\usepackage{makecell}
\usepackage{amsmath}
\usepackage{amsfonts}
\usepackage{multirow}
\usepackage[table]{xcolor}
\usepackage{pifont}
\newcommand{\cmark}{\ding{51}}
\newcommand{\xmark}{\ding{55}}
\definecolor{samergray}{HTML}{E4F4E8}
\usepackage{newfloat}
\usepackage{listings}
\DeclareCaptionStyle{ruled}{labelfont=normalfont,labelsep=colon,strut=off} 
\floatstyle{ruled}
\newfloat{listing}{tb}{lst}{}
\floatname{listing}{Listing}

\usepackage{booktabs}
\usepackage{multirow}
\newcommand{\best}[1]{\textbf{#1}}
\newcommand{\second}[1]{\underline{#1}}
\usepackage{subcaption}
\title{Do All Visual Tokens Matter Equally?\\
Object-Evidence Preserving Token Merging for Vision-Language Retrieval}
\author {
    Suhyeong Park\textsuperscript{\rm 1,\rm 2},
    Junha Jung\textsuperscript{\rm 1,\rm 2},
    Jungwoo Park\textsuperscript{\rm 1,\rm 2},
    Jaewoo Kang\textsuperscript{\rm 1,\rm 2}\corresponding
}
\affiliations {
    \textsuperscript{\rm 1}Korea University\\
    \textsuperscript{\rm 2}AIGEN Sciences Inc.\\
    \texttt{\{pshpulip22, goodjungjun,jungwoo-park,kangj\}@korea.ac.kr}
}

\begin{document}

\maketitle

\begin{abstract}
Multi-vector vision-language retrieval preserves fine-grained visual evidence through maximum-similarity late interaction, but dense image-side tokens make storage and scoring expensive. Existing token compression methods reduce this cost, yet they can remove or collapse object- and region-level evidence that future query tokens may need to select. We propose SaMer, an object-aware token merging framework that compresses image-side post-projector tokens into \(K\) representative centroids while preserving the original late-interaction interface. SaMer uses object annotations only during training as a merge prior to discourage cross-instance mixing, requires no ground-truth bounding boxes or detectors at inference time, and adapts only the shared projection layer with frozen vision and language backbones. With \(K=64\), SaMer removes more than 93\% of image-side tokens and reduces ColPali storage by \(16.09\times\), while improving R@1 on Flickr30K and MSCOCO. These gains arise because object-aware merging preserves query-selectable object evidence that pruning or feature-only pooling can remove or collapse. SaMer also outperforms compression baselines and shows stronger phrase-level grounding, suggesting that efficient multi-vector retrieval depends not only on reducing token count, but on preserving the evidence future query tokens need to select.
\end{abstract}

\begin{links}
    \link{Code}{https://github.com/dmis-lab/SaMer}
\end{links}

\section{Introduction}

Efficient multi-vector vision-language retrieval requires preserving the object-, attribute-, and relation-level evidence that query tokens can select under late interaction. Multi-vector retrieval has been widely adopted in text retrieval since ColBERT~\cite{i1,i2}, which performs maximum-similarity (MaxSim) matching between contextualized query and document token embeddings. Recent vision-language retrievers such as ColPali~\cite{i3} extend this paradigm to visual inputs by storing image-side patch embeddings and comparing them with query tokens through MaxSim. Similarly, ColQwen2~\cite{i3} builds on Qwen2-VL~\cite{i4}, whose visual tokenization supports flexible image representation. These models use MaxSim to compare each query token with all image-side tokens and select the most similar visual token as evidence, making retrieval sensitive to objects, attributes, and relations rather than only global image-text similarity~\cite{add9,add10}.

However, preserving token-level image representations makes retrieval expensive in both storage and scoring. A multi-vector retriever stores hundreds to over a thousand image-side token embeddings per image, and retrieval requires MaxSim comparisons between query tokens and all stored image tokens. As the index grows, both memory footprint and scoring latency increase rapidly, limiting large-scale image search and retrieval-augmented visual question answering. The issue is especially pronounced in patch-based visual representations, where each image produces many local tokens that must be stored and compared during MaxSim scoring, making patch-level embeddings a major bottleneck for multi-vector VLM retrievers~\cite{i5,add7,i6}.

One strategy is to compress visual tokens through pruning or merging \cite{add2,add3,add4,add6,r15}. Recent multimodal acceleration methods reduce visual tokens for large vision-language models \cite{add1, r10}. Yet retrieval compression differs from token reduction because patch importance is query-dependent. 
For the same image, one query may depend on a small object, another on an attribute, and a third on a relation. If compression removes or incorrectly merges phrase-relevant evidence, late interaction can no longer recover the token that should support the match. This risk is acute under feature-based merging, where similar patches from different instances of the same category can collapse into the same representation and lose the distinction on which a query depends~\cite{add_i1, add_i2}. For multi-vector retrieval, the goal is not simply to reduce redundant tokens, but to preserve object- and region-level evidence in a form that remains selectable under future MaxSim queries.

To this end, we propose \textbf{S}emantic-\textbf{a}ware \textbf{Mer}ging (SaMer), an object-aware token merging framework that preserves object- and region-level semantic evidence for future MaxSim queries. SaMer compresses post-projector visual tokens into \(K\) merged tokens using feature-spatial soft assignment, forming each representative as a normalized weighted centroid while preserving the original late-interaction interface. During training, object annotations serve not as an auxiliary grounding loss, but as a merge prior that guides which visual tokens should be grouped together. This discourages cross-instance merging, helping compressed representatives retain object-level evidence for later MaxSim retrieval. At inference, SaMer requires no ground-truth bounding boxes (bbox) or object detector and performs annotation-free feature-spatial merging while remaining pluggable into existing multi-vector retrievers. Finally, SaMer uses projection-only adaptation, freezing the vision encoder and language backbone while training only the shared projection layer to keep merged tokens compatible with MaxSim scoring.

Because SaMer is designed to preserve object-level structure under compression, retrieval accuracy alone may not reveal whether compressed tokens retain the phrase-level evidence used by late interaction. We therefore complement retrieval evaluation with grounding-oriented metrics that measure whether query-token relevance remains concentrated inside annotated phrase regions. With \(K=64\), SaMer removes more than 93\% of image-side tokens, reduces ColPali image-side storage by \(16.09\times\), and improves R@1 from 77.0 to 82.4 on Flickr30K and from 47.4 to 51.6 on MSCOCO. Grounding results further show that SaMer better preserves phrase-level visual evidence than pruning- and pooling-based compression baselines at the same token budget. These results support our view that the effectiveness of retrieval compression depends on whether object evidence remains accessible to query tokens after compression. Our contributions are:

\begin{itemize}
    \item We frame visual-token compression for late-interaction retrieval as an evidence-preservation problem and identify object-instance collapse as a key failure mode of feature-based merging.

    \item We propose SaMer, an object-aware token merging method that uses training-time object annotations as a merge prior to discourage cross-instance merging, without requiring annotations at inference time.
 
    \item We show that SaMer at \(K=64\) removes over 93\% of visual tokens and reduces ColPali storage by \(16.09\times\), while outperforming pruning- and pooling-based compression baselines under matched budget and adaptation with stronger phrase-level grounding.
\end{itemize}

\section{Related Works}
\label{app:related_work}

\subsection{Vision-language retrieval and late interaction}

Vision-language retrieval has been widely studied through contrastive image-text representation learning and multimodal pretraining, including ALIGN~\cite{r1}, ALBEF~\cite{r2}, BLIP~\cite{r3}, and BLIP-2~\cite{r4}. While these models provide strong image-text representations, they can underrepresent local evidence when retrieval depends on fine-grained objects, attributes, or regions. Late-interaction retrieval addresses this limitation by preserving token-level matching, as in COIL~\cite{r5}, SPLADE~\cite{r6, r7}, PLAID~\cite{r8}, WARP~\cite{r9}, and multi-vector vision-language retrievers such as ColPali and ColQwen2. However, storing and scoring many visual token embeddings per candidate introduces storage and scoring costs.

\subsection{Visual token reduction and retrieval-side compression}

Visual token reduction has been explored through pruning and merging methods such as LLaVA-PruMerge~\cite{r10}, LVPruning~\cite{r11}, VisionZip~\cite{r12}, ZipVL~\cite{r13}, LiteLVLM~\cite{r14}, and ToMe~\cite{r15}. These methods mainly reduce transformer computation inside generative LVLMs or vision encoders, whereas SaMer compresses post-projector retrieval embeddings stored in the multi-vector index and used by late-interaction scoring. Recent work has studied retrieval-side compression for multi-vector vision-language retrievers. Prior work compares pruning and merging strategies for ColPali and ColQwen2~\cite{add_r1}, while HPC-ColPali~\cite{add_r2} combines quantization, attention-guided pruning, and binary encoding. SAP~\cite{add_r3} prunes structural anchor patches, and H-Pool~\cite{add_r4} performs modality-agnostic fixed-budget pooling for multi-vector indexes. Unlike these methods, SaMer uses object annotations during adaptation to shape the merge assignment and discourage cross-instance mixing, while requiring no detector at inference time.

\begin{figure*}[t]
\centering
\includegraphics[width=\textwidth, keepaspectratio]{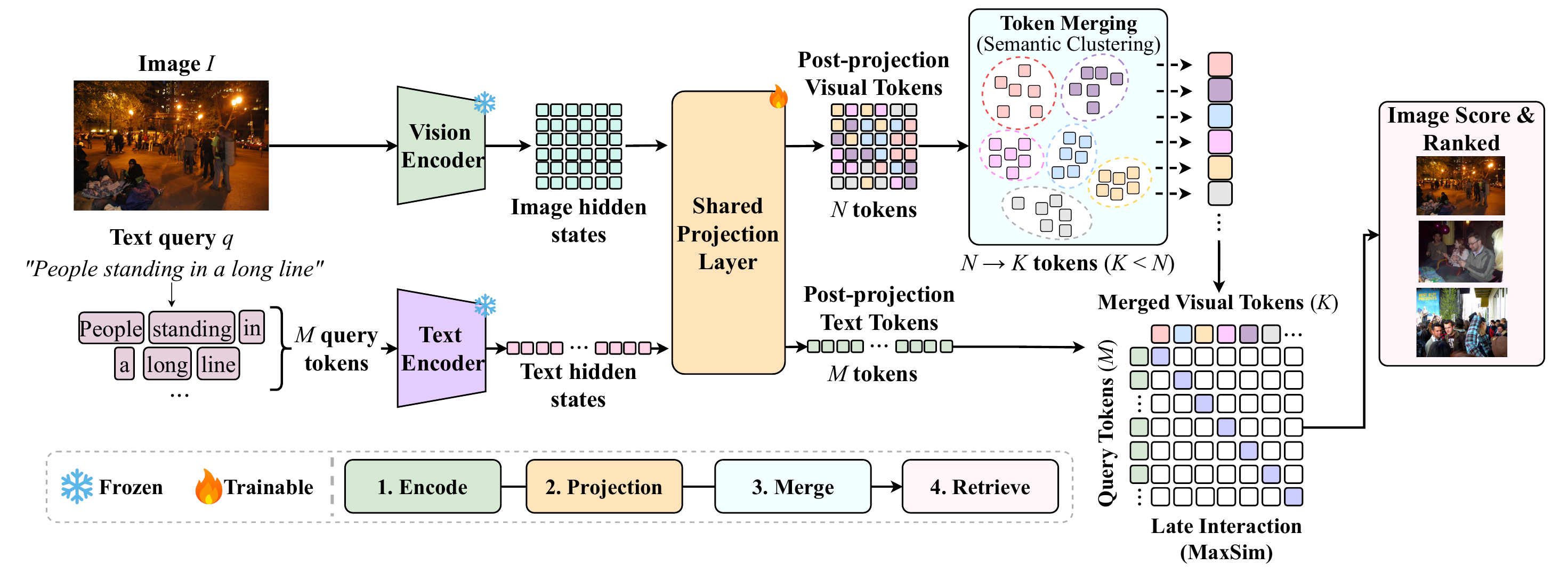}
\caption{
Overview of SaMer. Frozen vision and language encoders produce hidden states, a shared trainable projection layer maps both modalities into the retrieval space, and image-side post-projector tokens are compressed from \(N\) to \(K\) representative tokens before MaxSim scoring. During training, object-aware merging and projection-only adaptation encourage the compressed tokens to preserve object- and region-level evidence; at inference time, SaMer keeps the same late-interaction interface without requiring bbox or detectors.
}
\label{fig:overview}
\end{figure*}

\section{Preliminaries}

\subsection{Multi-vector Late Interaction}

Multi-vector vision-language retrieval represents images and text queries as sets of token embeddings and computes retrieval scores through late interaction. Given an image $I$, the vision encoder and projection layer produce $N$ visual tokens, denoted as $V(I)=\{v_i\}_{i=1}^{N}$, where $v_i\in\mathbb{R}^{D}$. Similarly, a text query $q$ is represented as $M$ query tokens $Q(q)=\{q_j\}_{j=1}^{M}$, where $q_j\in\mathbb{R}^{D}$. Both image and text tokens are projected into the same $D$-dimensional retrieval space.

The retrieval score is defined by MaxSim late interaction. Each query token is matched to the most similar visual token, and the query-image score is computed as
\begin{equation}
S(q,I) = \frac{1}{M}\sum_{j=1}^{M}\max_{i\in[N]} q_j^\top v_i .
\label{eq:maxsim}
\end{equation}

This scoring allows query tokens to match objects, attributes, or regions in the image, making multi-vector retrieval effective for fine-grained visual evidence. However, since every query token should be compared with every visual token, the number of visual tokens $N$ becomes the bottleneck for both storage and computation.

\subsection{Post-projector Token Compression}

In multi-vector vision-language retrieval, each image is represented by \(N\) post-projector visual tokens \(V=\{v_i\}_{i=1}^{N}\), where each token has been mapped into the shared image-text retrieval space. These tokens are stored in the retrieval index and used by MaxSim scoring. Therefore, full-token retrieval requires \(O(ND)\) storage per image for \(D\)-dimensional tokens and computes \(MN\) query-image token similarities per image, corresponding to \(O(MND)\) dot-product cost for a query with \(M\) tokens. Reducing the number of image-side post-projector tokens can directly lower both index storage and scoring cost while preserving the encoder and late-interaction retrieval interface. Unlike token pruning, which discards visual evidence before future queries are known, token merging aims to aggregate the token set into a smaller set of representative tokens.

\section{Methods}

Our framework has three steps, all designed to keep the original late-interaction interface while replacing the image side with compressed representatives. First, SaMer compresses the image-side post-projector tokens into \(K\) representative tokens using feature-spatial soft assignment. 
Second, during training only, object annotations guide which tokens may be merged by penalizing assignments that mix different object instances.
Third, SaMer adapts only the shared projection layer so that the compressed tokens remain compatible with MaxSim scoring, while the vision encoder and language backbone stay frozen. Figure~\ref{fig:overview} illustrates the overall process.

\subsection{Post-projector Visual Token Merging}
\label{sec:post_projector_merging}

SaMer reduces the number of visual tokens while keeping the late-interaction interface unchanged. Given the full post-projector visual token set \(V=\{v_i\}_{i=1}^{N}\), SaMer assigns these tokens to \(K\) representatives, where \(K \ll N\), and constructs one merged token for each representative. These post-projector embeddings are the vectors stored in the retrieval index and compared with query tokens by MaxSim. Therefore, compressing them directly reduces index storage and query-time similarity computation without modifying the vision encoder or the retrieval scoring function.

We construct the compressed representation with feature-spatial centroid merging. Each visual token \(v_i\) is associated with a spatial coordinate \(p_i \in [0,1]^2\), and each representative \(k\) maintains a feature centroid \(\mu_k\) and a spatial centroid \(s_k\) initialized from \(K\) uniformly spaced seed tokens in the flattened image-token grid. SaMer assigns tokens to representatives using a distance that combines feature similarity and spatial coherence:
\begin{equation}
d(i,k)
=
\underbrace{\left(1-v_i^\top \mu_k\right)}_{\text{feature similarity}}
+
\gamma
\underbrace{\left\|p_i-s_k\right\|_2^2}_{\text{spatial coherence}} ,
\label{eq:distance}
\end{equation}
where \(\gamma\) controls the strength of the spatial term. The feature term groups semantically similar tokens, while the spatial term discourages visually similar but spatially distant regions from being merged too aggressively.

SaMer converts this distance \(d(i,k)\) into soft assignment weights \(a_{i,k}\) and forms each representative as the corresponding weighted centroid with \(\ell_2\) normalization:
\begin{equation}
\begin{aligned}
a_{i,k}
&=
\mathrm{softmax}_{k}
\left(
-\frac{d(i,k)}{\tau_s}
\right), \\
r_k
&=
\mathrm{normalize}_2
\left(
\frac{
\sum_{i=1}^{N} a_{i,k}v_i
}{
\sum_{i=1}^{N} a_{i,k}
}
\right).
\end{aligned}
\label{eq:centroid_token}
\end{equation}
Here, \(a_{i,k}\) is the assignment weight from token \(i\) to representative \(k\), and \(\tau_s\) is the soft-assignment temperature. Unlike pruning, the centroid construction in Eq.~\ref{eq:centroid_token} does not discard visual tokens; instead, it aggregates them into compact representatives. The compressed representation \(R(I)=\{r_k\}_{k=1}^{K}\) is scored with the same late-interaction rule as Eq.~\ref{eq:maxsim}, except that MaxSim is taken over merged tokens:
\begin{equation}
S_K(q,I)
=
\frac{1}{M}
\sum_{j=1}^{M}
\max_{k\in[K]}
q_j^\top r_k .
\label{eq:compressed_maxsim}
\end{equation}
Thus, in the training-free setting, SaMer acts as a cache-time compression method where the query encoder, image encoder, projection layer, and scoring rule remain unchanged, and only the stored image-side representation is replaced with \(K\) merged tokens.

\subsection{Object-aware Merge Prior}
\label{sec:object_aware_prior}

When object-level annotations are available during training, SaMer uses them as a weak merge prior rather than as an auxiliary grounding loss. The goal is to make merged representatives more object-consistent, since visually similar regions may correspond to different object instances and a purely feature-based assignment may collapse them into the same representative. For each visual token \(v_i\), we assign a bbox label \(b_i\) using its spatial coordinate \(p_i\): tokens outside all bboxes receive a background label, tokens inside a bbox receive the corresponding object instance label, and tokens inside multiple boxes are assigned to the smallest bbox. These labels are used only during training to define the merge prior; background and context tokens are not removed from the compressed representation.

To build the object-aware prior, we first estimate which bbox labels dominate each representative. We do this with a stop-gradient hard assignment \(c_i=\arg\min_{k\in[K]} d(i,k)\), computed from the feature-spatial distance before applying the object-aware penalty. This hard assignment is not used to form the final merged token; it is used only to estimate the bbox-label distribution of each representative, so gradients do not backpropagate through \(c_i\):
\begin{equation}
\begin{gathered}
P_k(b)
=
\frac{
\sum_{i=1}^{N} \mathbf{1}[c_i=k]\mathbf{1}[b_i=b]
}{
\sum_{i=1}^{N} \mathbf{1}[c_i=k]+\epsilon
},
\\[0.35em]
P_{\mathrm{inst}}(i,k)
=
1-P_k(b_i).
\end{gathered}
\label{eq:instance_penalty}
\end{equation}
This penalty is small when token \(i\) shares the bbox label with most tokens in representative \(k\), and large when assigning \(i\) would mix evidence from different object instances.

After estimating this prior, SaMer returns to soft assignment to construct the actual training-time merged tokens. The object-aware penalty is added to the feature-spatial distance, so cross-instance assignments receive lower weight:
\begin{equation}
\begin{aligned}
a_{i,k}
&=
\mathrm{softmax}_{k}
\left(
-\frac{d(i,k)+P_{\mathrm{inst}}(i,k)}{\tau_s}
\right), \\[0.35em]
r_k^{\mathrm{soft}}
&=
\mathrm{normalize}_2
\left(
\frac{
\sum_{i=1}^{N} a_{i,k}v_i
}{
\sum_{i=1}^{N} a_{i,k}
}
\right).
\end{aligned}
\label{eq:object_soft_assignment_centroid}
\end{equation}
where the softmax is taken over representatives \(k=1,\ldots,K\) for each token \(i\). During training, SaMer uses centroids \(r_k^{\mathrm{soft}}\) in Eq.~\ref{eq:object_soft_assignment_centroid} for compressed MaxSim scoring. The penalty reshapes the soft assignment weights, through which gradients flow to the projected embeddings and merged centroids. At inference time, SaMer does not use bboxes or object labels. Instead, it performs bbox-free feature-spatial soft assignment, while the object-aware prior is reflected in the adapted projection space. We use soft assignment because it performs best empirically among centroid construction variants, which are compared in the Appendix.

\begin{table*}[t]
\centering
\footnotesize
\resizebox{\textwidth}{!}{
\begin{tabular}{@{}l*{9}{c}|*{3}{c}@{}}
\toprule
\multirow{2}{*}{Method}
& \multicolumn{3}{c}{Flickr30K}
& \multicolumn{3}{c}{MSCOCO}
& \multicolumn{3}{c|}{ImageCoDe}
& \multicolumn{3}{c}{DocVQA} \\
\cmidrule(lr){2-4}
\cmidrule(lr){5-7}
\cmidrule(lr){8-10}
\cmidrule(l){11-13}
& R@1 & R@5 & nDCG@10
& R@1 & R@5 & nDCG@10
& R@1 & R@5 & nDCG@10
& R@1 & R@5 & nDCG@10 \\
\midrule

\multicolumn{10}{@{}l|}{\textit{Single-vector}} & \multicolumn{3}{c@{}}{} \\
CLIP
& 58.2 & 82.9 & 73.7
& 30.8 & 56.0 & 47.7
& 1.4 & 7.0 & 5.7
& 6.7 & 13.8 & 11.8 \\
OpenCLIP
& 76.5 & 93.5 & 87.0
& 48.5 & 73.2 & 64.4
& 2.2 & 9.1 & 7.3
& \second{15.3} & 26.8 & \second{23.1} \\
MetaCLIP
& 62.8 & 85.8 & 77.2
& 35.7 & 61.7 & 52.8
& 2.1 & 7.6 & 6.3
& 5.4 & 14.4 & 11.5 \\
EVA-CLIP
& \best{79.9} & \second{94.7} & \best{89.2}
& 51.0 & 74.6 & 66.1
& \second{2.5} & 8.8 & \second{7.5}
& 12.9 & 25.1 & 19.8 \\
DFN-CLIP
& 79.6 & \best{94.9} & \second{89.0}
& \best{53.4} & \best{77.8} & \best{69.0}
& 1.9 & \best{9.8} & \best{7.7}
& 12.4 & \second{27.7} & 21.8 \\
SigLIP2
& \second{79.8} & 94.2 & 88.8
& \second{52.1} & \second{75.2} & \second{66.9}
& \best{2.6} & \second{9.4} & 7.5
& \best{15.5} & \best{29.1} & \best{24.3} \\

\midrule
\multicolumn{10}{@{}l|}{\textit{VLM-based}} & \multicolumn{3}{c@{}}{} \\
VLM2Vec-v2
& 41.7 & 68.8 & 59.0
& 20.2 & 44.1 & 36.5
& 1.0 & 5.1 & 4.4
& 4.0 & 10.4 & 8.5 \\
GME
& \best{74.6} & \best{91.8} & \best{85.2}
& \best{47.3} & \best{71.1} & \best{62.7}
& \best{2.3} & \best{9.1} & \best{7.7}
& \best{41.5} & \best{60.3} & \best{53.3} \\
VisRAG
& \second{69.6} & \second{90.0} & \second{82.2}
& \second{42.5} & \second{67.5} & \second{58.9}
& \second{1.7} & \second{7.0} & \second{6.0}
& \second{34.8} & \second{51.2} & \second{45.6} \\

\midrule
\multicolumn{10}{@{}l|}{\textit{Multi-vector}} & \multicolumn{3}{c@{}}{} \\

ColPali
& \second{77.0} & \second{92.9} & \second{86.9}
& \second{47.4} & \second{71.7} & \second{63.3}
& 5.4 & 16.0 & 13.2
& \best{51.0} & \best{66.5} & \best{60.7} \\
\quad HPC
& 54.0 & 77.8 & 69.0
& 30.6 & 52.5 & 45.4
& 3.6 & 11.3 & 9.0
& 26.4 & 39.7 & 34.9 \\
\quad H-Pool
& 73.7 & 91.4 & 84.7
& 44.9 & 69.6 & 60.9
& \second{5.6} & 15.6 & 12.9
& 45.0 & 61.9 & 55.5 \\
\quad SAP
& 68.3 & 88.6 & 80.8
& 39.6 & 63.7 & 55.4
& 3.8 & 12.5 & 10.1
& 40.4 & 55.2 & 50.2 \\
\rowcolor{samergray}
\quad SaMer (w/o FT)
& 73.5 & 91.1 & 84.5
& 45.1 & 69.7 & 61.1
& 5.2 & \best{17.0} & \second{14.1}
& 44.6 & \second{62.1} & 55.6 \\
\rowcolor{samergray}
\quad SaMer (w/ FT)
& \best{82.4} & \best{96.3} & \best{90.9}
& \best{51.6} & \best{75.5} & \best{67.1}
& \best{5.9} & \best{17.0} & \best{14.4}
& \second{45.7} & 60.5 & \second{55.6} \\

ColQwen2
& \second{73.6} & \second{91.3} & \second{84.7}
& \second{43.3} & \second{68.0} & \second{59.5}
& 4.7 & 14.6 & 12.2
& \best{53.2} & \best{67.9} & \best{63.0} \\
\quad HPC
& 67.1 & 88.7 & 80.4
& 40.1 & 64.8 & 56.5
& 4.4 & 13.9 & 11.6
& 39.9 & 55.2 & 49.6 \\
\quad H-Pool
& 71.8 & 90.8 & 83.6
& 42.2 & 67.2 & 58.6
& 4.7 & 14.2 & 12.3
& \second{50.1} & \second{64.8} & \second{59.3} \\
\quad SAP
& 69.3 & 89.9 & 82.2
& 40.7 & 65.8 & 57.2
& \second{4.8} & 13.7 & 11.8
& 45.5 & 59.4 & 54.7 \\
\rowcolor{samergray}
\quad SaMer (w/o FT)
& 71.3 & 90.4 & 83.3
& 42.1 & 67.0 & 58.4
& 4.8 & \second{15.6} & \best{13.0}
& 47.2 & 63.9 & 57.3 \\
\rowcolor{samergray}
\quad SaMer (w/ FT)
& \best{79.3} & \best{94.7} & \best{88.7}
& \best{47.5} & \best{72.3} & \best{63.5}
& \best{5.1} & \best{16.2} & \second{12.9}
& 48.8 & 62.5 & 58.1 \\
\bottomrule
\end{tabular}
}
\caption{
Retriever-only results across four benchmarks. SaMer compresses the image-side representation to \(K=64\) tokens and is compared with compression baselines applied to the same backbone when available. Shaded rows indicate SaMer variants. \textbf{Bold} and \underline{underline} indicate the best and second-best results within each method group or backbone family, respectively.
}
\label{tab:retrieval_only}
\end{table*}

\subsection{Compression-Aware Projection-Only Adaptation}
\label{sec:projection_adaptation}

Training-free SaMer can be directly applied to an existing multi-vector retriever, but small-\(K\) compression creates a mismatch between the representation used by the original retriever and the compressed representation used by SaMer. The projection layer is trained assuming that all \(N\) visual tokens are available for late-interaction scoring, whereas SaMer scores only \(K\) merged tokens. We therefore adapt only the image-text shared projection layer using the compressed score \(S_K\), while keeping the vision encoder and language backbone frozen. This aligns the retrieval space with compressed tokens without changing the backbone representations or introducing a new retrieval architecture.

During adaptation, token merging remains non-parametric: it introduces no learnable merge parameters. However, because the merged centroids are computed from projected visual embeddings through soft assignment and weighted averaging, gradients from the compressed MaxSim score still update the projection layer. Thus, the projection is learned to produce features that remain effective after merging. When object annotations are available, the object-aware prior shapes the soft assignments without adding an auxiliary loss, and SaMer is optimized only with a multi-positive InfoNCE retrieval loss~\cite{add_m1} over compressed MaxSim scores:
\begin{equation}
\mathcal{L}_{ret}
=
-\log
\frac{
\sum_{I^+\in\mathcal{P}(q)}
\exp(S_K(q,I^+)/\tau)
}{
\sum_{I\in\mathcal{B}}
\exp(S_K(q,I)/\tau)
}.
\label{eq:retrieval_loss}
\end{equation}
Here, \(\mathcal{P}(q)\) is the set of positive images for query \(q\), \(\mathcal{B}\) is the batch candidate set, and \(\tau\) is the temperature. At inference time, \(K\) merged visual tokens are cached, and retrieval uses the same late-interaction scoring rule over the compressed image representation.

\section{Experiments}

\subsection{Datasets}

We evaluate SaMer with retriever-only evaluation on Flickr30K, MSCOCO, ImageCoDe, and DocVQA~\cite{e4}. Flickr30K and MSCOCO evaluate natural-image text-to-image retrieval, where queries often refer to objects, attributes, and relations in natural scenes. Since SaMer is adapted only on the Flickr30K-Entities~\cite{e2} training split, we treat Flickr30K as the in-domain benchmark and MSCOCO as a cross-dataset benchmark. ImageCoDe~\cite{e5} evaluates compositional retrieval over visually similar images, where small object, attribute, or relation differences can determine the correct image. This makes ImageCoDe useful for testing whether compression preserves subtle query-dependent evidence among near-miss candidates. DocVQA provides a document-domain reference, where retrieval often depends on sparse OCR tokens and layout cues rather than natural-object evidence. It therefore serves as a boundary case for object-centric compression in a domain that requires broad textual and spatial coverage. We report Recall@1 (R@1), Recall@5 (R@5), and nDCG@10.

\subsection{Baselines}

We compare against single-vector, VLM-based, multi-vector, and retrieval-side compression baselines. Single-vector baselines include CLIP~\cite{e8}, OpenCLIP~\cite{e9}, MetaCLIP~\cite{e10}, EVA-CLIP~\cite{e11}, DFN-CLIP~\cite{e12}, and SigLIP2~\cite{e13}, which use global image-query embeddings. VLM-based baselines include VLM2Vec-v2~\cite{e14}, GME~\cite{e15}, and VisRAG~\cite{e16}, which use stronger multimodal backbones. Multi-vector baselines include ColPali and ColQwen2. Retrieval-side compression baselines include H-Pool, HPC, and SAP, which reduce image-side retrieval embeddings through pooling, attention-based pruning, or structural anchor pruning. We apply SaMer to the multi-vector retrievers in training-free and adapted settings. For controlled comparisons, compression baselines use the same backbones and \(K=64\) visual tokens when applicable, following their method-specific settings. Unless otherwise stated, SaMer uses \(K=64\), adapts the shared projection layer for 3 epochs on Flickr30K-Entities, and keeps the backbones frozen. Optimization details are in the Appendix.

\subsection{Results}

Table~\ref{tab:retrieval_only} presents retriever-only performance across natural-image, compositional-image, and document-image benchmarks. SaMer is most effective on natural-image retrieval, where redundant visual tokens coexist with query-specific object and region evidence. On Flickr30K, SaMer improves ColPali R@1 from 77.0 to 82.4 and ColQwen2 R@1 from 73.6 to 79.3, suggesting that compact object-aware representatives can improve retrieval under aggressive token reduction. It also outperforms H-Pool, the strongest external compression baseline, whose R@1 is 73.7 with ColPali and 71.8 with ColQwen2. The same trend holds on MSCOCO, where SaMer improves ColPali R@1 from 47.4 to 51.6 and outperforms H-Pool, SAP, and HPC. These results suggest that feature-similarity pooling alone may be less effective than object-aware merging with projection-only adaptation for preserving object evidence under MaxSim retrieval. Merging also tends to preserve retrieval evidence better than random or spatial pruning, as reported in the Appendix.

SaMer performs strongly on ImageCoDe, where visually similar images must be distinguished using subtle object, attribute, and relation cues. With ColPali, SaMer improves R@1 from 5.4 to 5.9 and nDCG@10 from 13.2 to 14.4, supporting our view that efficient compression should preserve query-selectable object evidence rather than only retain globally salient or visually similar patches. DocVQA serves as a boundary case beyond the object-centric retrieval setting targeted by SaMer: document images require broad OCR and layout coverage, so aggressive token compression can remove sparse textual evidence rather than redundant visual patches. SaMer is therefore not optimized for this setting, but remains competitive with compressed baselines using a compact image-side representation.

\section{Analysis of SaMer}

\subsection{Ablation Study}
\label{sec:ablation_study}

\paragraph{Is the Gain Only from Adaptation?}

\begin{table}[t]
\centering
\footnotesize
\renewcommand{\arraystretch}{1.06}
\setlength{\tabcolsep}{3.0pt}
\resizebox{\columnwidth}{!}{
\begin{tabular}{@{}l*{6}{c}@{}}
\toprule
\multirow{2}{*}{Method}
& \multicolumn{2}{c}{Flickr30K}
& \multicolumn{2}{c}{MSCOCO}
& \multicolumn{2}{c}{ImageCoDe} \\
\cmidrule(lr){2-3}
\cmidrule(lr){4-5}
\cmidrule(l){6-7}
& R@1 & nDCG@10
& R@1 & nDCG@10
& R@1 & nDCG@10 \\
\midrule

\multicolumn{7}{@{}l}{\textit{ColPali}} \\
\quad H-Pool
& 79.2 & 88.7
& \second{49.7} & \second{65.3}
& \best{5.9} & \second{13.8} \\
\quad HPC
& 67.7 & 81.1
& 37.8 & 53.4
& \second{4.7} & 11.8 \\
\quad SAP
& \second{80.2} & \second{89.5}
& 47.3 & 62.9
& 4.6 & 12.6 \\
\rowcolor{samergray}
\quad SaMer
& \best{82.4} & \best{90.9}
& \best{51.6} & \best{67.1}
& \best{5.9} & \best{14.4} \\

\addlinespace[0.25em]

\multicolumn{7}{@{}l}{\textit{ColQwen2}} \\
\quad H-Pool
& 77.0 & 87.0
& 46.1 & 62.1
& \second{5.0} & \second{12.9} \\
\quad HPC
& 77.9 & 87.8
& 45.3 & 61.7
& 4.9 & 12.7 \\
\quad SAP
& \second{78.2} & \second{88.2}
& \second{46.4} & \second{62.5}
& 5.0 & 12.4 \\
\rowcolor{samergray}
\quad SaMer
& \best{79.3} & \best{88.7}
& \best{47.5} & \best{63.5}
& \best{5.1} & \best{12.9} \\
\bottomrule
\end{tabular}
}
\caption{
Adapted compression results at \(K=64\) under projection-only adaptation. Shaded rows indicate SaMer; \textbf{bold} and \underline{underline} denote the best and second-best results.
}
\label{tab:adapted_compression_ablation}
\end{table}

One possible explanation for SaMer's gains is that projection-only adaptation alone improves the compressed retrieval space. To separate adaptation from the merge design, Table~\ref{tab:adapted_compression_ablation} compares SaMer with other compression methods trained under the same adaptation setting. All methods use the same frozen backbone, the same projection-only update, and the same \(K=64\) visual token budget.

SaMer achieves the best R@1 and nDCG@10 on natural-image and compositional retrieval metrics for ColPali and ColQwen2. This suggests that the improvement is not a fine-tuning effect. If projection-only adaptation were the sole explanation, we would expect the gap between SaMer and the other adapted compression variants to be smaller under the same training setup. Instead, the results suggest that SaMer remains stronger because its merge rule better controls which visual evidence survives compression. By discouraging tokens from different object instances from being mixed, SaMer preserves object- and region-level evidence more effectively than pruning, pooling, or ratio-based compression.

\begin{figure}[t]
\centering
\includegraphics[width=\columnwidth]{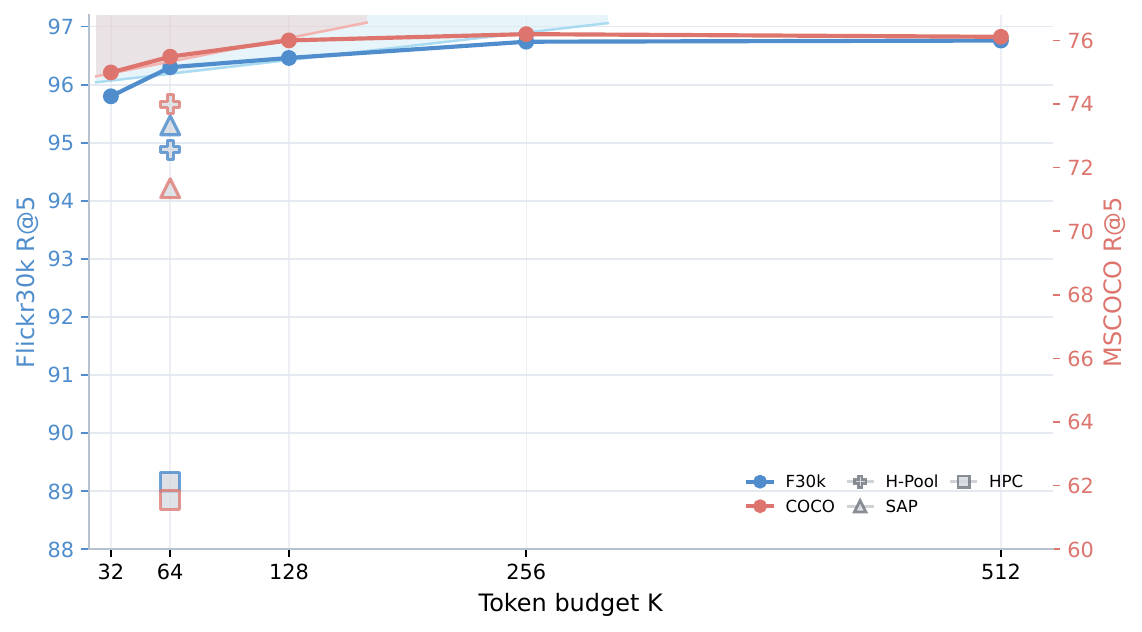}
\caption{
Token budget and R@5 trade-off. Solid lines show SaMer across token budgets, with baselines marked at \(K=64\). SaMer shows diminishing returns beyond \(K=64\) and outperforms baselines under the same budget.
}
\label{fig:token_budget_pareto}
\end{figure}

\paragraph{Compression Budget Analysis.}
Table~\ref{tab:budget} reports retrieval performance as the token budget \(K\) varies, and Figure~\ref{fig:token_budget_pareto} visualizes the trade-off between token budget and R@5. Retrieval performance improves as \(K\) increases from 32 to 64, but the gain becomes small beyond that point. On Flickr30K, R@5 rises from 95.8 at \(K=32\) to 96.3 at \(K=64\), and then increases only to 96.8 at \(K=512\). MSCOCO shows the same pattern, increasing from 75.0 at \(K=32\) to 75.5 at \(K=64\), and reaching 76.1 at \(K=512\). This suggests that a small set of merged tokens is sufficient to preserve most retrieval performance on these benchmarks.

Figure~\ref{fig:token_budget_pareto} compares SaMer with competing compression methods at the same \(K=64\) budget. Under this controlled setting, SaMer lies above H-Pool, SAP, and HPC on both Flickr30K and MSCOCO, indicating a better quality-budget trade-off at the same token budget. Taken together, these results make \(K=64\) a favorable operating point, balancing strong retrieval quality with substantial image-side compression. We use \(K=64\) throughout the main experiments.

\begin{table}[t]
\centering
\footnotesize
\renewcommand{\arraystretch}{1.06}
\setlength{\tabcolsep}{3.0pt}
\resizebox{\columnwidth}{!}{
\begin{tabular}{@{}lc*{6}{c}@{}}
\toprule
\multirow{2}{*}{\makecell{$K$\\variant}}
& \multirow{2}{*}{\shortstack{Comp.\\ratio (\%)}}
& \multicolumn{3}{c}{Flickr30K}
& \multicolumn{3}{c}{MSCOCO} \\
\cmidrule(lr){3-5}
\cmidrule(l){6-8}
& & R@1 & R@5 & nDCG@10
& R@1 & R@5 & nDCG@10 \\
\midrule
512
& 50.29
& \best{83.7} & \best{96.8} & \best{91.7}
& 52.2 & \second{76.1} & 67.5 \\
256
& 75.15
& \second{83.5} & \second{96.7} & \second{91.6}
& \best{52.4} & \best{76.2} & \best{67.7} \\
128
& 87.57
& 83.1 & 96.5 & 91.3
& \second{52.3} & 76.0 & \second{67.6} \\
\rowcolor{samergray}
64 (ours)
& 93.79
& 82.4 & 96.3 & 90.9
& 51.6 & 75.5 & 67.1 \\
32
& 96.89
& 81.3 & 95.8 & 90.3
& 51.3 & 75.0 & 66.4 \\
\bottomrule
\end{tabular}
}
\caption{
Compression budget analysis on Flickr30K and MSCOCO as \(K\) varies. \(K=64\) provides a favorable trade-off between retrieval quality and image-side compression.
}
\label{tab:budget}
\end{table}

\paragraph{Merge Component Study.}
Table~\ref{tab:merge_component_compact} analyzes the contribution of each merge component. Feature-only merging groups visually similar tokens, but adding spatial coherence alone provides little benefit. Flickr30K R@1 changes from 80.7 to 80.4, and MSCOCO R@1 remains nearly unchanged. This suggests that local smoothness is not enough for retrieval-oriented compression because query-relevant evidence can depend on object identity rather than spatial proximity alone. In other words, spatial coherence helps organize nearby tokens, but it does not by itself make the merged representation object-aware.

The object-aware merge prior provides the main gain. It raises Flickr30K R@1 to 82.4 and MSCOCO R@1 to 51.6, while also improving grounding quality. This suggests that the gain comes not only from projection adaptation, but also from constraining feature-spatial merging to respect object-instance structure. We evaluate grounding with BoxMass, RegionHit, and CoverageIoU, which measure whether phrase relevance remains aligned with annotated object regions. 
The metric definitions are provided in the Appendix.
BoxMass increases from 47.8 to 54.2 and CoverageIoU from 13.1 to 16.4, indicating that the object-aware prior improves retrieval by preserving phrase-level visual evidence inside the correct object regions. 
Full merge-component metrics are reported in the Appendix.

\begin{table}[t]
\centering
\footnotesize
\renewcommand{\arraystretch}{1.08}
\setlength{\tabcolsep}{3.4pt}
\resizebox{\columnwidth}{!}{
\begin{tabular}{@{}lccc|cccc@{}}
\toprule
\multirow{2}{*}{Variant}
& \multicolumn{3}{c|}{Components}
& \multicolumn{2}{c}{Retrieval}
& \multicolumn{2}{c}{Grounding} \\
\cmidrule(lr){2-4}
\cmidrule(lr){5-6}
\cmidrule(l){7-8}
& \makecell{Feat.} & \makecell{Spat.} & \makecell{Obj.}
& \makecell{Flickr30K\\R@1} & \makecell{MSCOCO\\R@1}
& BoxMass & CovIoU \\
\midrule
(A)
& \cmark & \xmark & \xmark
& 80.7 & 49.8 
& 47.6 & 13.0 \\
(B)
& \cmark & \cmark & \xmark
& 80.4 & 49.8
& 47.8 & 13.1 \\
\rowcolor{samergray}
SaMer
& \cmark & \cmark & \cmark
& \textbf{82.4} & \textbf{51.6}
& \textbf{54.2} & \textbf{16.4} \\
\bottomrule
\end{tabular}
}
\caption{
SaMer combines feature, spatial, and object-aware components.
}
\label{tab:merge_component_compact}
\end{table}

\subsection{Grounding Comparison}
\label{sec:grounding_comparison}

\begin{figure}[t]
\centering
\includegraphics[width=\columnwidth]{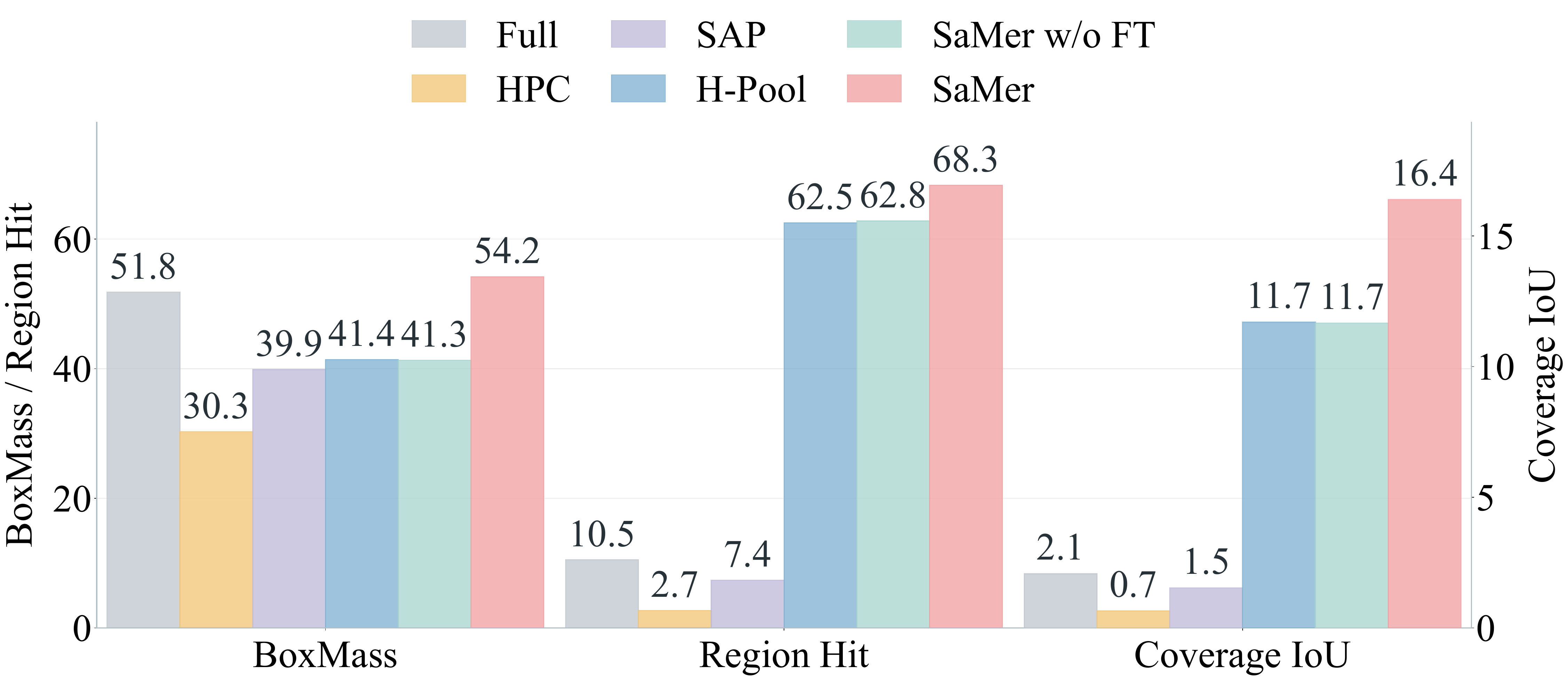}
\caption{
Grounding comparison between full ColPali, compression baselines, and SaMer variants. BoxMass and RegionHit use the left axis, while CoverageIoU uses the right axis. SaMer improves localization and evidence concentration under compression.
}
\label{fig:grounding_overall}
\end{figure}

Figure~\ref{fig:grounding_overall} evaluates whether compressed representations preserve phrase-level visual evidence. Compared with the full ColPali representation, compression baselines show mixed behavior: some improve weak localization, but their relevance can remain spread across nearby objects or background regions. This suggests that reducing visual tokens can still leave enough signal to reach the general target area, but does not necessarily preserve object-specific evidence for precise phrase grounding. SaMer shows a more consistent pattern. Training-free SaMer improves RegionHit from 10.5 to 62.8 and CoverageIoU from 2.1 to 11.7, indicating that feature-spatial merging alone helps the compressed representation retain spatial access to the correct phrase region even without projection adaptation.

Object-aware projection-only adaptation further improves relevance concentration. SaMer raises BoxMass from 41.3 to 54.2, surpassing the full-token baseline of 51.8, while improving RegionHit to 68.3 and CoverageIoU to 16.4. This distinction matters because a method may reach the target region while still assigning substantial relevance to surrounding context, nearby objects, or visually similar background patches. The BoxMass gain shows that SaMer does not merely make the relevance map overlap with the target bbox; it concentrates more phrase-specific evidence inside the annotated object region. With the improvements in RegionHit and CoverageIoU, this suggests that the object-aware prior helps merged tokens preserve both localization and evidence concentration under aggressive compression. Qualitative grounding examples are provided in the Appendix.

\subsection{Efficiency Analysis}

Table~\ref{tab:efficiency_summary} summarizes the storage and scoring efficiency gains of SaMer under the same late-interaction retrieval interface. Since SaMer compresses only the cached image-side representation, storage is reported per one million cached images, while MaxSim comparisons and QPS are measured by scoring all queries against all candidate images in each evaluation dataset. For ColPali, which stores 1,030 visual tokens per image before compression, SaMer reduces storage by \(16.09\times\) and MaxSim comparisons by \(16.1\times\) on both Flickr30K and MSCOCO. This reduction translates into substantial throughput gains, increasing QPS by \(4.3\times\) on Flickr30K and \(9.1\times\) on MSCOCO.

For ColQwen2, the full representation contains fewer visual tokens, so the gains are smaller but still meaningful. SaMer reduces storage by \(3.73\times\) on Flickr30K and \(1.99\times\) on MSCOCO, reduces MaxSim comparisons by \(3.7\times\) and \(2.0\times\), and improves QPS by \(1.9\times\) and \(3.1\times\), respectively. Throughput gains are affected by memory movement and batching, so they do not scale one-to-one with comparison reduction, but the improvements remain substantial. These results show that SaMer reduces both storage and retrieval latency, lowering scoring cost without changing the query encoder or MaxSim scoring rule. Storage and latency details appear in the Appendix.

\begin{table}[t]
\centering
\small
\renewcommand{\arraystretch}{1.15}
\setlength{\tabcolsep}{3pt}
\resizebox{\columnwidth}{!}{
\begin{tabular}{@{}lccc ccc@{}}
\toprule
\multirow{2}{*}{Variant}
& \multicolumn{3}{c}{Flickr30K}
& \multicolumn{3}{c}{MSCOCO} \\
\cmidrule(lr){2-4}
\cmidrule(l){5-7}
& \makecell{Storage/\\1M (GB)} & \makecell{MaxSim\\Ops (\(10^{12}\))} & QPS
& \makecell{Storage/\\1M (GB)} & \makecell{MaxSim\\Ops (\(10^{12}\))} & QPS \\
\midrule
\multicolumn{7}{l}{\textit{ColPali}} \\
Full
& 263.7 & 36.87 & 625.5
& 263.7 & 845.51 & 154.2 \\
SaMer
& 16.4 {\scriptsize(\(16.09\times\))}
& 2.29 {\scriptsize(\(16.1\times\))}
& 2716.3 {\scriptsize(\(4.3\times\))}
& 16.4 {\scriptsize(\(16.09\times\))}
& 52.54 {\scriptsize(\(16.1\times\))}
& 1406.7 {\scriptsize(\(9.1\times\))} \\
\midrule
\multicolumn{7}{l}{\textit{ColQwen2}} \\
Full
& 61.0 & 8.01 & 1381.6
& 32.6 & 97.50 & 314.7 \\
SaMer
& 16.4 {\scriptsize(\(3.73\times\))}
& 2.15 {\scriptsize(\(3.7\times\))}
& 2674.7 {\scriptsize(\(1.9\times\))}
& 16.4 {\scriptsize(\(1.99\times\))}
& 48.99 {\scriptsize(\(2.0\times\))}
& 969.7 {\scriptsize(\(3.1\times\))} \\
\bottomrule
\end{tabular}
}
\caption{
SaMer reduces image-side storage and MaxSim operations while improving query throughput under the same late-interaction interface. Parentheses indicate improvement over the corresponding full multi-vector retriever.
}
\label{tab:efficiency_summary}
\end{table}

\section{Conclusion}

In this work, we introduced SaMer, an object-aware token merging framework for efficient multi-vector vision-language retrieval. Motivated by the view that image-side compression should preserve query-selectable evidence rather than merely reduce token count, SaMer compresses post-projector image tokens into \(K\) representative centroids. By using object annotations only during training, SaMer discourages cross-instance merging while requiring no bboxes or object detectors at inference time. With projection-only adaptation, SaMer aligns compressed visual representations with the existing retrieval space while keeping the vision and language backbones frozen. Experiments show that \(K=64\) merged tokens substantially reduce storage and MaxSim scoring cost while maintaining strong retrieval performance and phrase-level grounding. Overall, our results suggest that efficient late-interaction retrieval depends not only on token count, but also on preserving query-selectable evidence.

\section*{Acknowledgments}
We thank Taewhoo Lee, Hyeon Hwang, and Yein Park for their insightful feedback and discussions. This research was supported by the National Research Foundation of Korea (NRF-2023R1A2C3004176), the Ministry of Health \& Welfare, Republic of Korea (HR20C002103), the Ministry of Science and ICT through Seoul National University Hospital (RS-2023-00262002), the National Research Foundation of Korea (NRF) grant funded by the Korea government (MSIT and MOE) (No. RS-2025-16652968), the ICT Creative Consilience Program through the Institute of Information \& Communications Technology Planning \& Evaluation (IITP) grant funded by the Korea government (MSIT) (IITP-2026-RS-2020-II201819), the Ministry of Science and ICT (MSIT) and National IT Industry Promotion Agency (NIPA) through the Hyperscale AI Healthcare Service Support Program (H0701-25-1003), and grants of the Korea Health Technology R\&D Project through the Korea Health Industry Development Institute (KHIDI), funded by the Ministry of Health \& Welfare, Republic of Korea (RS-2025-25462758 and RS-2022-KH129295).

\bibliography{aaai2027}

\clearpage
\appendix

\section{Centroid Construction Variants}
\label{app:centroid_variants}

\begin{table}[t]
\centering
\footnotesize
\renewcommand{\arraystretch}{1.06}
\setlength{\tabcolsep}{3.2pt}
\resizebox{\columnwidth}{!}{
\begin{tabular}{@{}lccc ccc@{}}
\toprule
\multirow{2}{*}{Centroid variant}
& \multicolumn{3}{c}{Flickr30K}
& \multicolumn{3}{c}{MSCOCO} \\
\cmidrule(lr){2-4}
\cmidrule(l){5-7}
& R@1 & R@5 & nDCG@10
& R@1 & R@5 & nDCG@10 \\
\midrule
Mean
& 80.3 & 95.4 & 89.7
& 49.3 & 73.9 & 65.3 \\
Normalized
& 82.3 & 96.1 & 90.7
& 51.5 & 75.5 & 67.0 \\
Medoid
& 80.6 & 95.6 & 89.9
& 50.1 & 74.6 & 65.9 \\
Object weighted
& 82.2 & 96.1 & 90.7
& 51.6 & 75.4 & 67.0 \\
\rowcolor{samergray}
Soft assignment (ours)
& \textbf{82.4} & \textbf{96.3} & \textbf{90.9}
& \textbf{51.6} & \textbf{75.5} & \textbf{67.1} \\
\bottomrule
\end{tabular}
}
\caption{
Centroid construction ablation on Flickr30K and MSCOCO. Soft assignment, used by SaMer, achieves the best performance across all metrics, showing that differentiable soft assignment centroids better preserve retrieval evidence than mean, normalized, medoid, or object-weighted alternatives.
}
\label{tab:centroid_variant}
\end{table}

Table~\ref{tab:centroid_variant} compares different ways of constructing merged representative tokens. The mean variant averages assigned tokens without additional normalization, while the normalized variant applies normalization after centroid construction. The medoid variant selects the token closest to the cluster center, preserving an existing token rather than forming an averaged representative. The object-weighted variant gives higher weight to tokens associated with object regions during centroid construction.

Soft assignment, used by SaMer, performs best across all metrics on both Flickr30K and MSCOCO. Compared with hard mean or medoid-style representatives, soft assignment centroids better preserve retrieval-relevant evidence by allowing tokens to contribute to clusters according to assignment confidence. The gain over the object-weighted variant further suggests that object information is most effective when used to shape the merge assignment itself, rather than only reweighting tokens after clusters have already been formed.

\section{Implementation Details}
\label{app:implementation_details}

We implement SaMer on top of existing multi-vector vision-language retrievers by modifying only the image-side post-projector representation. Unless otherwise stated, all main experiments use $K=64$ merged image tokens, a spatial weight of $\gamma=0.1$, a retrieval temperature of $\tau=0.07$, and a soft-assignment temperature of $\tau_s=0.07$. During compression-aware adaptation, the vision encoder and language backbone are frozen, and only the shared projection layer is updated. We train for 1,746 optimization steps using AdamW with a per-GPU batch size of 32, gradient accumulation over 2 steps, and an effective batch size of 256 across 4 NVIDIA A100 80GB GPUs. We use a learning rate of $2\times10^{-4}$, a weight decay of $1\times10^{-5}$, and a cosine learning-rate schedule with a warmup ratio of 0.1.

Object annotations are used only during training to construct the object-aware merge prior. Each image token is assigned a label based on its spatial coordinate: tokens outside all bounding boxes are treated as background, tokens inside a single bounding box receive the corresponding object-instance label, and tokens contained in multiple bounding boxes are assigned to the smallest box. These labels are used solely to compute the instance inconsistency penalty and are not treated as an auxiliary supervision signal. At inference time, SaMer discards all object labels and bounding boxes and performs annotation-free feature-spatial soft assignment.

\section{Grounding Metrics}
\label{app:grounding_metrics}

We define grounding metrics to test whether compressed tokens retain phrase-level visual evidence. Given a query phrase, we compute a normalized relevance map \(\hat{S}(x)\in[0,1]\) over image/grid locations \(x\), and compare it with the ground-truth phrase box mask \(G(x)\in\{0,1\}\). Since SaMer aims to preserve object- and region-level evidence under compression, the metrics evaluate relevance concentration, weak localization, and spatial coverage.

\paragraph{BoxMass.}
BoxMass is our primary grounding metric. It measures whether phrase relevance is concentrated inside the annotated object region, following attention correctness and activation-based grounding evaluations~\cite{a1,a2}:
\begin{equation}
\mathrm{BoxMass}
=
\frac{
\sum_x \hat{S}(x)G(x)
}{
\sum_x \hat{S}(x)+\epsilon
}.
\label{eq:boxmass_app}
\end{equation}
A higher BoxMass means that more phrase-specific evidence falls inside the correct object region, directly testing whether merging preserves the queried visual evidence.

\paragraph{RegionHit.}
RegionHit measures whether the model localizes the target phrase at all, following hit-based grounding and pointing-style protocols~\cite{a4,a3}. Let \(T=\{0.1,0.2,\ldots,0.9\}\) and \(M_t(x)=\mathbf{1}[\hat{S}(x)\geq t]\) denote the thresholded high-relevance region. We define
\begin{equation}
\mathrm{RegionHit}
=
\frac{1}{|T|}
\sum_{t\in T}
\mathbf{1}
\left[
\mathrm{IoU}(M_t,G)\geq 0.05
\right].
\label{eq:regionhit_app}
\end{equation}
We use a low IoU threshold because the goal is weak phrase-region localization rather than segmentation-quality prediction.

\paragraph{CoverageIoU.}
CoverageIoU measures how well the high-relevance region covers the target box, following activation-map localization analyses~\cite{a5,a6}:
\begin{equation}
\mathrm{CoverageIoU}
=
\frac{1}{|T|}
\sum_{t\in T}
\mathrm{IoU}(M_t,G).
\label{eq:coverage_iou_app}
\end{equation}
Unlike RegionHit, CoverageIoU measures the degree of spatial agreement rather than only whether the relevance map reaches the target. All three metrics are higher-is-better and together evaluate whether SaMer preserves phrase-level evidence after aggressive visual token compression.

\section{Merge Component Study}
\label{app:full_merge_component}

Table~\ref{tab:full_merge_component} reports the merge component study. The compact table in the main paper reports representative retrieval and grounding metrics, while this table includes all retrieval metrics on Flickr30K and MSCOCO and all object-consistency metrics. The full results show the same trend as the main paper: the spatial term alone does not substantially improve over feature-only merging, whereas the object-aware penalty consistently improves retrieval and grounding.

\section{Merging vs. Pruning Comparison}
\label{app:merge_vs_pruning}

\begin{table}[t]
\centering
\footnotesize
\renewcommand{\arraystretch}{1.06}
\setlength{\tabcolsep}{4.0pt}
\resizebox{\columnwidth}{!}{
\begin{tabular}{@{}l*{6}{c}@{}}
\toprule
\multirow{2}{*}{Variant}
& \multicolumn{3}{c}{Flickr30K}
& \multicolumn{3}{c}{MSCOCO} \\
\cmidrule(lr){2-4}
\cmidrule(l){5-7}
& R@1 & R@5 & nDCG@10
& R@1 & R@5 & nDCG@10 \\
\midrule
ColPali
& 77.0 & 92.9 & 86.9
& 47.4 & 71.7 & 63.3 \\
\midrule
\multicolumn{7}{l}{\textit{w/o FT}} \\
Random pruning
& 68.6 & 88.5 & 80.9
& 40.5 & 65.3 & 56.7 \\
Spatial pruning
& 68.8 & 88.6 & 81.3
& 41.1 & 65.8 & 57.3 \\
\rowcolor{samergray}
SaMer
& \textbf{73.5} & \textbf{91.1} & \textbf{84.5}
& \textbf{45.1} & \textbf{69.7} & \textbf{61.1} \\
\midrule
\multicolumn{7}{l}{\textit{w/ FT}} \\
Random pruning
& 79.9 & 94.9 & 89.0
& 48.1 & 72.7 & 64.0 \\
Spatial pruning
& 79.5 & 94.9 & 89.0
& 49.0 & 73.2 & 64.7 \\
\rowcolor{samergray}
SaMer
& \textbf{82.4} & \textbf{96.3} & \textbf{90.9}
& \textbf{51.6} & \textbf{75.5} & \textbf{67.1} \\
\bottomrule
\end{tabular}
}
\caption{
Merging vs. pruning on Flickr30K and MSCOCO under training-free and adapted settings. SaMer consistently outperforms pruning baselines.
}
\label{tab:merge_vs_prune}
\end{table}

Table~\ref{tab:merge_vs_prune} shows that merging is more effective than pruning for multi-vector retrieval. In the training-free setting, SaMer incurs a smaller drop than random and spatial pruning, improving Flickr30K R@1 from 68.6 and 68.8 to 73.5, and MSCOCO R@1 from 40.5 and 41.1 to 45.1. This gap reflects a key difference between the two strategies: pruning removes visual evidence before the query is known, whereas merging keeps it in aggregated representatives. After projection-only adaptation, SaMer remains strongest across all metrics, reaching 82.4 R@1 on Flickr30K and 51.6 on MSCOCO. These results show that representative centroids preserve retrieval-relevant visual evidence better than discarding tokens, and that the advantage persists after adapting the projection space.

\section{Qualitative Grounding Examples}
\label{app:qualitative_grounding}

\begin{figure*}[t]
\centering
\begin{subfigure}[t]{\textwidth}
    \centering
    \includegraphics[width=\textwidth]{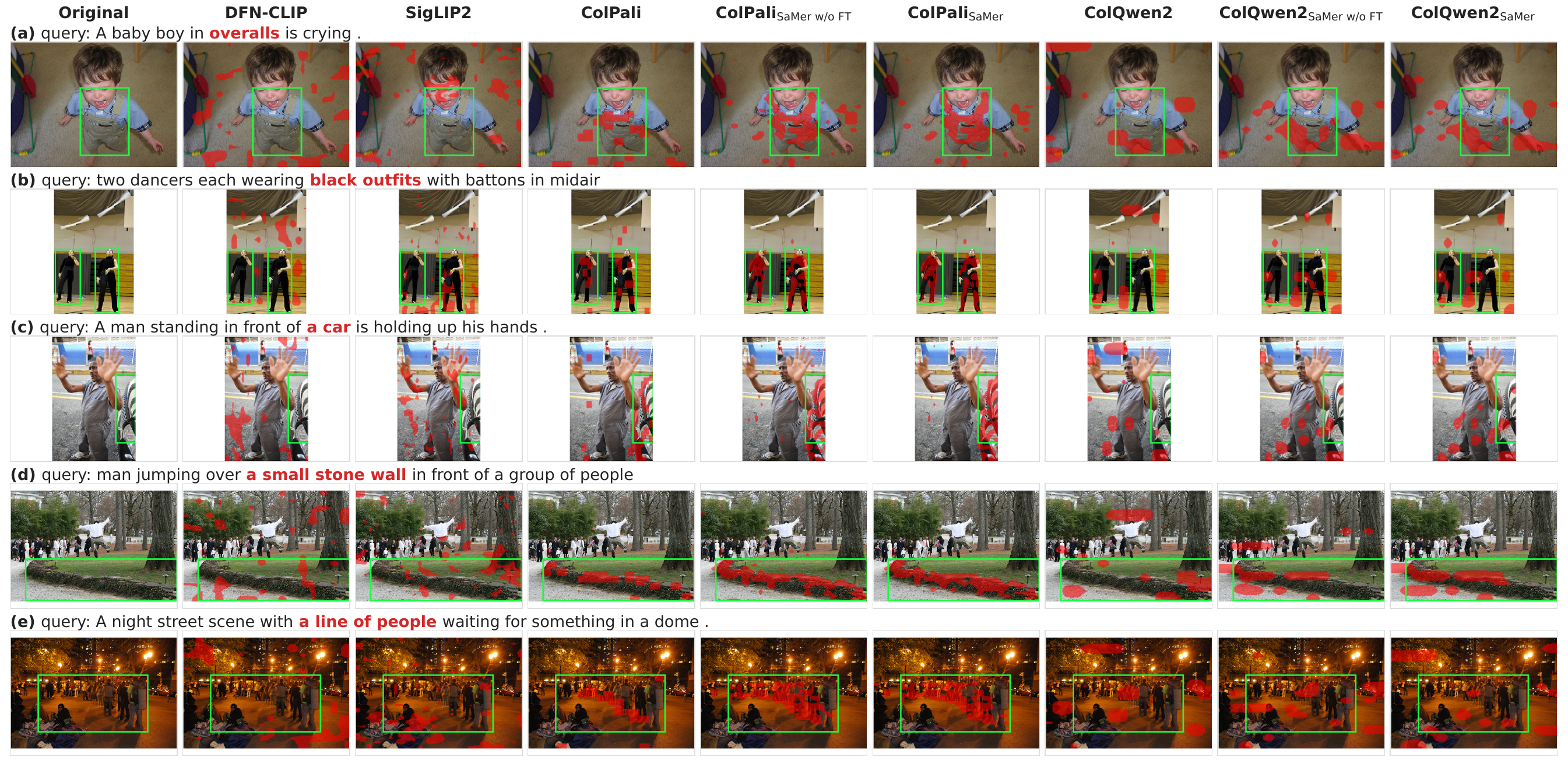}
    \caption{Comparison with single-vector baselines, full multi-vector retrievers, and SaMer variants.}
    \label{fig:visualization_example_main}
\end{subfigure}

\vspace{0.5em}

\begin{subfigure}[t]{\textwidth}
    \centering
    \includegraphics[width=\textwidth]{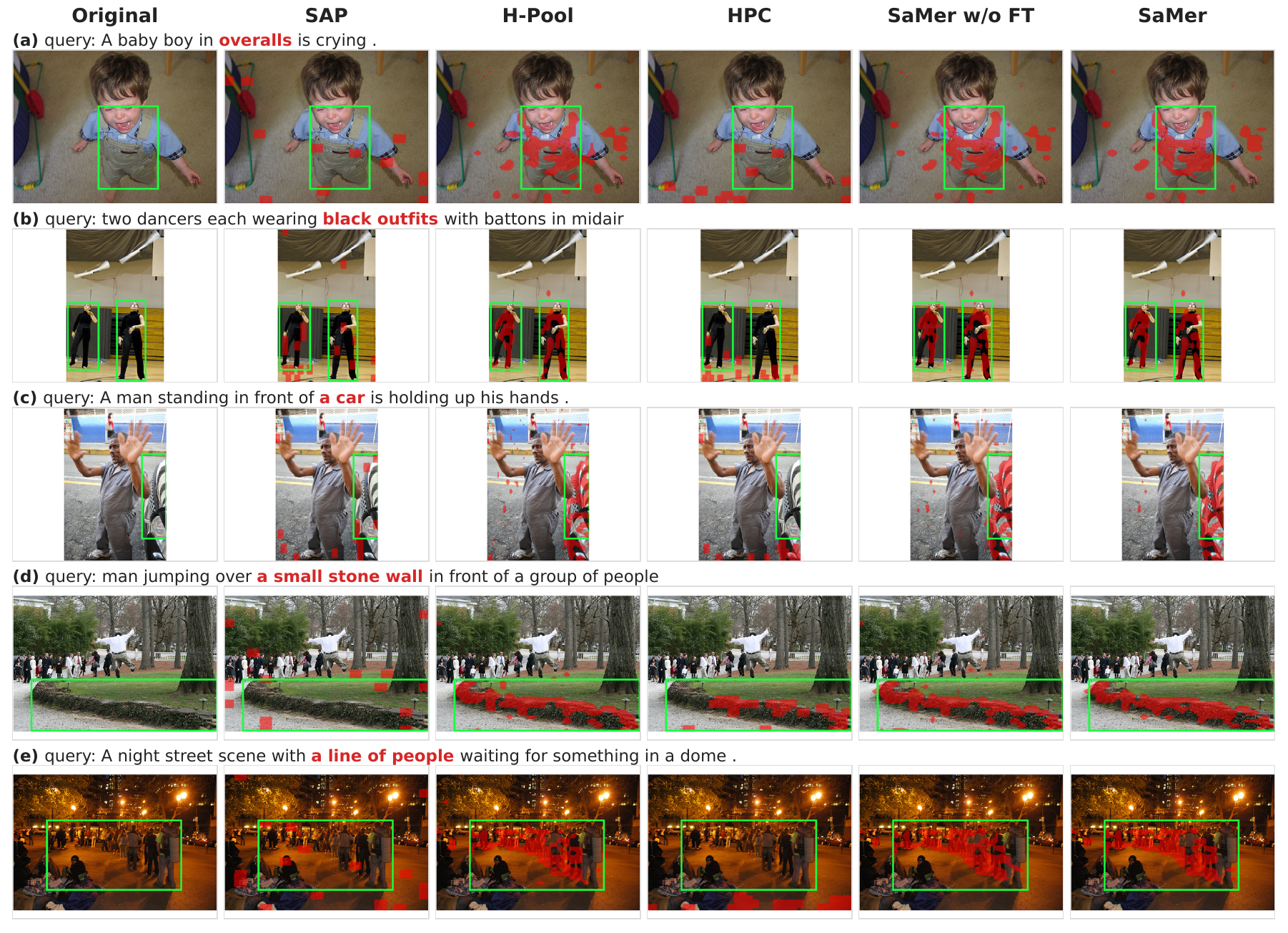}
    \caption{Comparison with compression methods applied to multi-vector retrievers.}
    \label{fig:visualization_example_compression}
\end{subfigure}

\caption{
Qualitative grounding examples. Red denotes the phrase-relevance map and green denotes the ground-truth box for the highlighted phrase. 
(a) compares SaMer with single-vector baselines and full multi-vector retrievers. 
(b) compares SaMer with compression methods applied to multi-vector retrievers. 
Across both settings, SaMer better concentrates phrase relevance inside the annotated region, while compression baselines often spread relevance to nearby objects or background regions.
}
\label{fig:visualization_examples}
\end{figure*}

Figure~\ref{fig:visualization_examples} provides qualitative grounding examples. Figure~\ref{fig:visualization_example_main} compares single-vector baselines, full multi-vector retrievers, and SaMer variants, while Figure~\ref{fig:visualization_example_compression} compares compression methods applied to multi-vector retrievers. The compression comparison shows that baseline methods often recover part of the target region but can spread relevance to nearby objects or background regions, such as surrounding clothing, road areas, or non-target people. SaMer more consistently concentrates relevance inside the annotated phrase region after compression, supporting the quantitative grounding results.

\section{Efficiency Measurements}
\label{app:detailed_efficiency}

\begin{table}[t]
\centering
\footnotesize
\renewcommand{\arraystretch}{1.06}
\setlength{\tabcolsep}{3.4pt}
\resizebox{\columnwidth}{!}{
\begin{tabular}{@{}lccc ccc@{}}
\toprule
\multirow{2}{*}{Variant}
& \multicolumn{3}{c}{Flickr30K}
& \multicolumn{3}{c}{MSCOCO} \\
\cmidrule(lr){2-4}
\cmidrule(l){5-7}
& Tokens/img & Storage/1M & Reduction
& Tokens/img & Storage/1M & Reduction \\
\midrule
\multicolumn{7}{l}{\textit{ColPali}} \\
Full
& 1030.0 & 263.7 GB & \(1.00\times\)
& 1030.0 & 263.7 GB & \(1.00\times\) \\
\rowcolor{samergray}
SaMer
& 64.0 & 16.4 GB & \(16.09\times\)
& 64.0 & 16.4 GB & \(16.09\times\) \\
\midrule
\multicolumn{7}{l}{\textit{ColQwen2}} \\
Full
& 238.4 & 61.0 GB & \(1.00\times\)
& 127.4 & 32.6 GB & \(1.00\times\) \\
\rowcolor{samergray}
SaMer
& 64.0 & 16.4 GB & \(3.73\times\)
& 64.0 & 16.4 GB & \(1.99\times\) \\
\bottomrule
\end{tabular}
}
\caption{
Detailed image-side embedding storage for one million images under FP16 storage with embedding dimension \(D=128\).
}
\label{tab:detailed_storage}
\end{table}

\begin{table}[t]
\centering
\footnotesize
\renewcommand{\arraystretch}{1.06}
\setlength{\tabcolsep}{2.5pt}
\resizebox{\columnwidth}{!}{
\begin{tabular}{@{}lccc ccc@{}}
\toprule
\multirow{2}{*}{Variant}
& \multicolumn{3}{c}{Flickr30K}
& \multicolumn{3}{c}{MSCOCO} \\
\cmidrule(lr){2-4}
\cmidrule(l){5-7}
& \makecell{MaxSim\\Ops (\(10^{12}\))} & Latency(s) & QPS
& \makecell{MaxSim\\Ops (\(10^{12}\))} & Latency(s) & QPS \\
\midrule
\multicolumn{7}{l}{\textit{ColPali}} \\
Full
& 36.87 & 7.994 & 625.5
& 845.51 & 160.879 & 154.2 \\
\rowcolor{samergray}
SaMer
& 2.29 {\scriptsize(\(16.1\times\))}
& 1.841 {\scriptsize(\(4.3\times\))}
& 2716.3 {\scriptsize(\(4.3\times\))}
& 52.54 {\scriptsize(\(16.1\times\))}
& 17.636 {\scriptsize(\(9.1\times\))}
& 1406.7 {\scriptsize(\(9.1\times\))} \\
\midrule
\multicolumn{7}{l}{\textit{ColQwen2}} \\
Full
& 8.01 & 3.619 & 1381.6
& 97.50 & 78.837 & 314.7 \\
\rowcolor{samergray}
SaMer
& 2.15 {\scriptsize(\(3.7\times\))}
& 1.869 {\scriptsize(\(1.9\times\))}
& 2674.7 {\scriptsize(\(1.9\times\))}
& 48.99 {\scriptsize(\(2.0\times\))}
& 25.585 {\scriptsize(\(3.1\times\))}
& 969.7 {\scriptsize(\(3.1\times\))} \\
\bottomrule
\end{tabular}
}
\caption{
Detailed retrieval efficiency comparison. MaxSim Ops denotes total token-level similarity comparisons; parentheses indicate improvement over the corresponding full multi-vector retriever.
}
\label{tab:detailed_efficiency}
\end{table}

Table~\ref{tab:detailed_storage} reports the raw storage estimates used to compute the storage reduction ratios in the main paper. Storage is estimated for one million images under FP16 storage with embedding dimension \(D=128\). Table~\ref{tab:detailed_efficiency} reports the corresponding scoring measurements, including total MaxSim operations, latency, and query throughput. These detailed measurements support the efficiency summary in the main paper and show that reducing visual tokens lowers both the theoretical MaxSim comparison count and the measured retrieval latency.

\section{SaMer Algorithm}
\label{app:algorithm}

Algorithm~\ref{alg:samer_training} summarizes the compression-aware adaptation procedure of SaMer. The key difference between training and inference is the object-aware prior. During training, box labels define an instance inconsistency penalty inside the soft assignment. During inference, the object prior is discarded and the image-side representation is constructed using bbox-free feature-spatial soft assignment.

\begin{algorithm*}[t]
\caption{Compression-aware Adaptation and Inference with SaMer}
\begin{algorithmic}[1]
\REQUIRE Training batch $\mathcal{B}$, token budget $K$,
spatial weight $\gamma$, retrieval temperature $\tau$,
soft-assignment temperature $\tau_s$
\REQUIRE Frozen vision encoder \(f_v\), frozen language backbone \(f_t\), trainable shared projection layer \(g_\theta\)

\STATE \textbf{Training: compression-aware projection-only adaptation}
\FOR{each image-text pair \((I,q)\in\mathcal{B}\)}
    \STATE Extract hidden states \(H_I=f_v(I)\) and \(H_q=f_t(q)\)
    \STATE Project image and query tokens into the retrieval space:
    \[
    V=\{v_i\}_{i=1}^{N}=g_\theta(H_I), \qquad
    Q=\{q_j\}_{j=1}^{M}=g_\theta(H_q)
    \]
    \STATE Initialize \(K\) representatives with feature centroids \(\mu_k\) and spatial centroids \(s_k\)
    \STATE Assign each image token a training-time box label \(b_i\) using its spatial coordinate \(p_i\)
    \STATE Compute the feature-spatial distance:
    \[
    d(i,k)=\left(1-v_i^\top \mu_k\right)+\gamma\lVert p_i-s_k\rVert_2^2
    \]
    \STATE Compute stop-gradient hard representative assignment:
    \[
    c_i=\arg\min_{k\in[K]} d(i,k)
    \]
    \STATE Compute hard-assignment label distribution \(P_k(b)\) and instance penalty:
    \[
    P_{\mathrm{inst}}(i,k)=1-P_k(b_i)
    \]
    \STATE Compute object-aware soft assignment:
    \[
    a_{i,k}^{\mathrm{obj}}
    =
    \mathrm{softmax}_{k}
    \left(
    -\frac{d(i,k)+P_{\mathrm{inst}}(i,k)}{\tau_s}
    \right)
    \]
    \STATE Construct merged image tokens:
    \[
    r_k=
    \mathrm{normalize}_2
    \left(
    \frac{\sum_i a_{i,k}^{\mathrm{obj}}v_i}{\sum_i a_{i,k}^{\mathrm{obj}}}
    \right)
    \]
    \STATE Compute compressed late-interaction score:
    \[
    S_K(q,I)=\frac{1}{M}\sum_{j=1}^{M}\max_{k\in[K]}q_j^\top r_k
    \]
\ENDFOR
\STATE Update only \(g_\theta\) using the multi-positive retrieval loss

\vspace{0.4em}
\STATE \textbf{Inference: annotation-free retrieval}
\STATE Extract and project visual tokens \(V=\{v_i\}_{i=1}^{N}\)
\STATE Compress \(V\) into \(K\) merged tokens \(R(I)=\{r_k\}_{k=1}^{K}\) using bbox-free feature-spatial soft assignment
\STATE Cache \(R(I)\) in the retrieval index
\STATE Extract and project query tokens \(Q(q)=\{q_j\}_{j=1}^{M}\)
\STATE Score each query-image pair using compressed late interaction:
\[
S_K(q,I)=\frac{1}{M}\sum_{j=1}^{M}\max_{k\in[K]}q_j^\top r_k
\]
\RETURN Retrieval score \(S_K(q,I)\)
\end{algorithmic}
\label{alg:samer_training}
\end{algorithm*}

\begin{table*}[t]
\centering
\footnotesize
\resizebox{\textwidth}{!}{
\begin{tabular}{@{}lccc|ccc ccc|ccc@{}}
\toprule
\multirow{2}{*}{Variant}
& \multicolumn{3}{c|}{Merge Components}
& \multicolumn{3}{c}{Flickr30K}
& \multicolumn{3}{c|}{MSCOCO}
& \multicolumn{3}{c}{Grounding} \\
\cmidrule(lr){2-4}
\cmidrule(lr){5-7}
\cmidrule(lr){8-10}
\cmidrule(l){11-13}
& Feature & Spatial & Obj.\ prior
& R@1 & R@5 & nDCG@10
& R@1 & R@5 & nDCG@10
& BoxMass & RegionHit & Coverage IoU \\
\midrule
(A)
& \cmark & \xmark & \xmark
& 80.7 & 95.6 & 89.8
& 49.8 & 73.7 & 65.4
& 47.6 & 67.9 & 13.0 \\
(B)
& \cmark & \cmark & \xmark
& 80.4 & 95.5 & 89.7
& 49.8 & 73.9 & 65.5
& 47.8 & 68.0 & 13.1 \\
\rowcolor{samergray}
SaMer
& \cmark & \cmark & \cmark
& \textbf{82.4} & \textbf{96.3} & \textbf{90.9}
& \textbf{51.6} & \textbf{75.5} & \textbf{67.1}
& \textbf{54.2} & \textbf{68.3} & \textbf{16.4} \\
\bottomrule
\end{tabular}
}
\caption{
The object-aware merge prior improves both retrieval performance and object consistency across all reported metrics.
}
\label{tab:full_merge_component}
\end{table*}

\end{document}